\documentclass[acmsmall]{acmart}

\AtBeginDocument{%
  \providecommand\BibTeX{{%
    \normalfont B\kern-0.5em{\scshape i\kern-0.25em b}\kern-0.8em\TeX}}}

\setcopyright{acmcopyright}
\acmJournal{TKDD}
\acmYear{2022}  \acmPrice{15.00}\acmDOI{10.1145/3494563}

\acmJournal{TKDD}
\acmVolume{16}
\acmNumber{4}
\acmArticle{78}
\acmMonth{1}

\usepackage{bbm}
\newcommand{\1}{\mathbbm{1}}

\newtheorem{remark}{Remark}

\usepackage{enumitem}

\usepackage{amsmath}

\usepackage[ruled]{algorithm2e} 

\DeclareMathOperator*{\argmax}{arg\,max}

\renewcommand{\C}{\mathcal{C}}
\newcommand{\A}{\mathcal{A}}
\renewcommand{\P}{\mathrm{P}}

\usepackage{color}

\begin{document}

\title{When Less Is More: Systematic Analysis of Cascade-Based Community Detection}

\author{Liudmila Prokhorenkova}
\email{ostroumova-la@yandex.ru}
\affiliation{%
  \institution{Yandex Research, MIPT, HSE University}
  \country{Russia}
}

\author{Alexey Tikhonov}
\email{altsoph@gmail.com}
\affiliation{%
  \institution{Yandex}
  \country{Germany}
}

\author{Nelly Litvak}
\email{n.litvak@utwente.nl}
\affiliation{%
  \institution{University of Twente, Eindhoven University of Technology}
  \country{the Netherlands}
}


\begin{abstract}
  Information diffusion, spreading of infectious diseases, and spreading of rumors are fundamental processes occurring in real-life networks. In many practical cases, one can observe when nodes become infected, but the underlying network, over which a contagion or information propagates, is hidden. Inferring properties of the underlying network is important since these properties can be used for constraining infections, forecasting, viral marketing, and so on. Moreover, for many applications, it is sufficient to recover only coarse high-level properties of this network rather than all its edges. This article conducts a systematic and extensive analysis of the following problem: Given only the infection times, find communities of highly interconnected nodes. This task significantly differs from the well-studied community detection problem since we do not observe a graph to be clustered. We carry out a thorough comparison between existing and new approaches on several large datasets and cover methodological challenges specific to this problem. One of the main conclusions is that the most stable performance and the most significant improvement on the current state-of-the-art are achieved by our proposed simple heuristic approaches agnostic to a particular graph structure and epidemic model. We also show that some well-known community detection algorithms can be enhanced by including edge weights based on the cascade data.
\end{abstract}



\maketitle

\section{Introduction}

Many social, biological, and information systems can be represented by networks whose nodes are items, and links are relations between them. In recent years, there has been a great deal of interest in analyzing propagation processes arising on such network structures. Examples of such processes are spreading of infectious diseases~\citep{hufnagel2004forecast} and computer viruses~\citep{wang2000computer}, promotion of products via viral marketing~\citep{kempe2003maximizing},  propagation of information~\citep{romero2011differences}, opinion dynamics~\citep{noorazar2020recent}, and so on. Usually, a contagion appears at some network node and then spreads to other nodes over the network's edges. In many practical scenarios, one can observe when nodes become infected, but the underlying network is hidden. For instance, during an epidemic, a person becomes ill but cannot tell who infected them; in viral marketing, we observe when customers buy products but not who influenced their decisions. Another example is Twitter, which sells a stream of tweets to other companies, but the social graph is not disclosed; for each retweet, only information about the cascade's root node is available. The hidden network inference problem received much attention recently~\citep{daneshmand2014estimating,rodriguez2014uncovering,rodriguez2012submodular,ramezani2017dani}. While these articles aim at recovering the actual network connections, such detailed information may not be needed in many cases.  Moreover, the unobserved network may not be unique or well defined: Users may communicate via different social networks or offline. In many practical applications, only some global properties of the network are important. For example, in viral marketing, one may wish to find the most influential users, while for recommendation systems, one may look for groups of users with similar preferences. 

In this article, we analyze the problem, which recently attracted interest in the literature  \citep{barbieri2017efficient,ramezani2018community}, of inferring {\it community structure} of a given network based solely on {\it cascades} (e.g., information or epidemic) propagating through this network. Communities are groups of highly interconnected nodes with relatively few edges joining nodes of different groups, such as groups by interests in social networks or pages on related topics on the Web~\citep{fortunato2010community}. Discovering communities is one of the most prominent tasks in network analysis with many recent applications, for example, in anomaly detection~\cite{ma2021comprehensive},  financial systems~\cite{kukreti2020perspective}, and dynamic brain networks~\cite{martinet2020robust}.
Compared to the traditional community detection, our work is quite different because we do not have the network available to us, we have only cascade data observed on this network. For each cascade, we observe only infected nodes and their infection timestamps. Leveraging cascade information for community detection is important, for example, for preventing the spreading of fake news~\cite{zhou2019network} and viral infections such as SARS-CoV-2~\citep{gosgens2021trade}.

The main contribution of this article is a thorough theoretical and empirical analysis of a developing area of inferring community structure based on cascade data, as stated formally in Section~\ref{sec:problem}. Specifically:
\begin{itemize}
\item We propose and analyze several new algorithms and compare them to the state-of-the-art. In particular, we consider two types of approaches: based on likelihood maximization under specific model assumptions and based on the clustering of a surrogate network (Section~\ref{sec:algorithms}). 
\item We propose new evaluation benchmarks (real and synthetic graphs with natural and synthetic cascades) and discuss important methodological questions on properly evaluating and comparing algorithms (Section~\ref{sec:setup}).
\item Via extensive experiments on a variety of networks,\footnote{The source code is available at \url{https://github.com/ostroumova-la/cascade_communities}.} we conclude that the most stable performance is obtained by our proposed heuristics that are agnostic to a particular graph structure and epidemic model. These heuristics outperform more advanced approaches and work equally well on different networks and epidemics of different types (Section~\ref{sec:experiments}).
\end{itemize}

Thus, our work significantly extends previous research on the subject: We analyze a comprehensive list of algorithms, we compare them empirically on several datasets with different cascade types, and we thoroughly discuss critical methodological aspects of the problem at hand, such as dealing with unobserved nodes, that have been overlooked in the literature.

A preliminary version of this article was presented in~\citet{prokhorenkova2019learning}. The current article substantially extends the previous work in several important directions:
(1)~significantly extending and improving the experimental part by adding more datasets and more cascade models; most importantly, we add the Twitter data with real cascades and natural communities (Section~\ref{sec:datasets}); (2) adding a new surrogate-graph-based method called CosineSim~(Section~\ref{sec:cosinesim}) to support our claim that comparing advanced methods with simple solutions is essential;
(3) analyzing the problem of how to aggregate and compare results on several datasets of different properties and sizes (Section~\ref{sec:average}), which is important for a systematic analysis on various data;
(4) discussing how to measure the quality of an algorithm, especially when the network is only partially observed~(Section~\ref{sec:metrics})~--- a very important question that may affect research conclusions; and (5) also investigating how to properly generate epidemic data for experiments (in case of not having real cascades) (Section~\ref{sec:cascades}). To sum up, this extended version provides a thorough analysis of the problem of community detection based on cascade data and discusses some methodological questions related to this task. We believe that our work would stimulate further theoretical and practical research in this area.

\section{Related Work}\label{sec:related_work}

In this section, we first discuss two topics that are closely related to our work and have been extensively studied in the literature: inferring the edges of the underlying network based on cascades and community detection in known graphs. Next, we address the state-of-the-art research on community detection from cascades, which we use as baselines.

\subsection{Network Inference from Cascades}

A series of recent articles addressed the following task: by observing infection (activation) times of nodes in several cascades, infer the underlying network's edges. \textsc{NetInf} algorithm developed  by~\citet{gomez2010inferring} is based on the maximization of the likelihood of the observed cascades under the assumption of a specific epidemic model. To make optimization feasible, for each cascade, \textsc{NetInf} considers only the most likely propagation tree. This algorithm was later improved by \textsc{MultiTree}~\citep{rodriguez2012submodular}, which includes all directed trees in the optimization and has better performance if the number of cascades is small. \textsc{NetRate} algorithm~\citep{rodriguez2011uncovering} infers not only edges but also infection rates. \textsc{NetRate} builds on an  epidemic model that is more tractable for theoretical analysis. (We describe this model in Section~\ref{sec:cascade_models}.) For this model, the likelihood optimization problem turns out to be convex.  \textsc{ConNIe} algorithm~\citep{myers2010convexity} also uses convex optimization, with the goal of inferring transmission probabilities for all edges. In~\citep{gomez2013structure,rodriguez2014uncovering}, it is additionally assumed that the underlying network is not static, and the proposed algorithm \textsc{InfoPath} provides estimates of the structure and temporal dynamics of the hidden network. \textsc{KernelCascade}~\citep{du2012learning} also extends \textsc{NetRate}, where the authors use a less restrictive cascade model. The DANI algorithm~\citep{ramezani2017dani} is interesting because it explicitly accounts for the community structure to enhance the inference of networks' edges. There are some other recent network inference algorithms not covered here, e.g.,~\cite{hoffmann2020learning} studies the methods for recovering a mixture of graphs and~\cite{gray2020bayesian} proposes a new Bayesian
method; see also references therein.

\subsection{Community Detection}\label{sec:community_detection}

Another direction closely related to the current research is community detection. For an overview of classic results in this area, we refer the reader to several survey articles~\citep{chakraborty2017metrics,coscia2011classification,fortunato2010community,fortunato2016community}. Many community detection algorithms are based on optimizing modularity~\citep{newman2004finding}, which measures the goodness of a partition for a given graph.
Probably the most well-known and widely used greedy modularity optimization algorithm is Louvain~\citep{blondel2008fast}, which iteratively moves nodes or groups of nodes from one community to another while improving modularity. Some of the methods discussed in this article use the Louvain algorithm.

In recent years, deep learning approaches became very popular for graph analysis~\citep{wu2020comprehensive}. Graph neural networks (GNNs) are especially effective for node classification and graph classification tasks. The main advantage of GNNs is their ability to naturally combine node features with graph structure. GNNs have also been applied to community detection~\citep{su2021comprehensive}. Relevant to our research are works on \emph{unsupervised} community detection~\citep{lobov2019unsupervised}. Often, such approaches optimize \emph{soft modularity} loss function. However, in the absence of node features, graph neural networks are often not superior to classic approaches.

Community detection algorithms most relevant to the current research are based on statistical inference. Such methods work as follows: we assume some random graph model parameterized by community assignments of nodes; then, for a given partition, we can compute the probability that this model generated the observed network structure (i.e., likelihood); finally, a partition providing the largest likelihood is chosen. In the literature, several possible choices of the random graph model and likelihood optimization methods were proposed. Early articles on likelihood optimization assumed stochastic block model (SBM) or planted partition model (PPM), where the probability that two nodes are connected depends only on their community assignments. 
Recently, a degree-corrected variant of this model was proposed~\citep{karrer2011stochastic}, and community detection methods based on this model were shown to have a better quality~\citep{newman2016community,peixoto2013parsimonious}. In a recent article~\citep{prokhorenkova2018community}, the independent Lancichinetti–Fortunato–Radicchi (ILFR) model was proposed and analyzed. It was shown that this model gives the best fit to a variety of real-world networks in terms of maximizing the likelihood. Additionally,~\citep{prokhorenkova2018community} extended the Louvain algorithm for optimizing any likelihood function. 

Closer to the current work, \citep{barbieri2013cascade, he2021discovering, sattari2018cascade} proposed to enhance community detection algorithms by using the information about spreading cascades. However, to find clusters, the authors use both the graph and the cascades. In contrast, the crucial assumption in our work is that the graph is not observed; hence these methods cannot be applied. Another similar setting is considered in~\citep{sanders2020clustering}. Here, it is assumed that we observe a trajectory of a  Markov chain spreading on a hidden stochastic block structure; the aim is to infer the clusters. Although the motivation is similar, our task is more complicated in two aspects: (1) In~\citep{sanders2020clustering}, the transitions of the Markov chain are defined solely by the membership in the communities, while in our work, we assume a hidden graph structure, which significantly complicates the problem. (2) We consider cascades instead of chains and, in particular, we do not know who infected whom. Recent work \cite{xing2021detecting} studies analytically the problem of detecting two communities from interactions in a gossip model, where some agents never change their state. In this work, we also leverage the (binary) states of the nodes, but the gossiping protocol fundamentally differs from cascades, and the assumption of two communities is too restrictive for our purposes. Finally,  \cite{hoffmann2020community} develops a Bayesian model for community detection based on time series observed at each of the network's nodes, without recovering the edges. However, their time series have a very different structure from the cascade data used in the current article.

\subsection{Community Inference from Cascades}

To the best of our knowledge, the article \citep{barbieri2013influence} extended in \citep{barbieri2017efficient} for the first time addressed the problem of community detection from cascades. In \citep{barbieri2013influence, barbieri2017efficient}, the cascade model includes the influence of individual nodes, and a membership level of a node in each community is inferred using the maximum-likelihood approach. The authors propose two algorithms: \textsc{C-IC}, which considers only participation of a node in a cascade, and \textsc{C-Rate}, which includes the time stamps but limits the node's influence by its community. Recently, \citep{ramezani2018community} proposed an alternative maximum-likelihood approach, which exploits the Markov property of the cascades. As an input, similarity scores of node pairs are computed based on their joint participation in cascades. The \textsc{R-CoDi} algorithm in \citep{ramezani2018community} starts with a random partition, while \textsc{D-CoDi} starts with a partition obtained by DANI~\citep{ramezani2017dani}. We use all four mentioned algorithms as our baselines. A recent work~\cite{suzuki2021effects} addressed a follow-up problem of cascade-based community detection when the cascade data are incomplete. They conclude that hiding information on the nodes that participate in many cascades considerably reduces the quality of community detection, while hiding random or least active nodes has much smaller effects.

A recent work by~\citet{peixoto2019network} proposes to simultaneously recover edges and communities from epidemic data and shows that the detection of communities significantly increases the accuracy of graph reconstruction. Another research by~\citet{chen2014community} also uses cascade data to identify (overlapping) communities. However, in contrast to the current work, they focus solely on Facebook and use more input data: post
information, personal information, interpersonal relations, and so on. 

\section{Problem setup}\label{sec:problem}

\subsection{General Setup}

\subsubsection{Cascades} We observe a set of cascades $\C = \{C_1, \ldots, C_r\}$ that propagate on a latent undirected network $G = (V,E)$ with $|V| = n$ nodes and $|E| = m$ edges. Each cascade $C \in \C$ is a record of observed node activation times, i.e., $C = \{(v_i, t_{v_{i}}^C)\}_{i=1}^{n_C}$, where $v_i$ is a node, $t_{v_i}^C$ is its activation time in $C$, $|C| = n_C$ is a size of a cascade. Note that we do not observe who infected whom.

\subsubsection{Communities}

We assume that $G$ is partitioned into non-overlapping communities: $\A = \{A_1, \ldots, A_k\}$, $\cup_{i = 1}^k A_i = V$, $A_i \cap A_j = \emptyset$ for $i \neq j$. We expect to observe high intra-community density of edges compared to inter-community density. In our experiments, the ground truth partitions $\A$ are available for all datasets (see Section~\ref{sec:datasets}). By observing only a set of cascades $\C$ we want to find a partition $\A'$ similar to $\A$. 

\subsection{Cascade Models}\label{sec:cascade_models}

\subsubsection{SIR Model}\label{sec:SIR}

The primary model for our experiments is a well-known Susceptible-Infected-Recovered (SIR) model~\citep{keeling2005networks}. Each node in the network can be in one of the three states:  susceptible, infected, or recovered. An infected node infects its susceptible neighbors with a rate $\alpha$, and the infection is spread simultaneously and independently along all edges. An infected node recovers with a rate $\beta$ and then stops spreading infection in the network. For each cascade $C$, we sample its infection rate $\alpha_C$ from the Lomax (shifted Pareto) distribution to model a variety of cascades: There can be minor or widely circulated news, small-scale epidemics or a global outbreak, and so on. The source node of a cascade is chosen uniformly at random.

\subsubsection{SI Model with Bounded Duration (SI-BD)}\label{sec:simple_SIR}
In some cases, the SIR model might not be tractable for theoretical analysis, so we assume a simpler diffusion model introduced in~\citep{rodriguez2011uncovering}. In this model, just as in the SIR model, an activated node infects its neighbors after an exponentially distributed time with intensity $\alpha$, but there is no recovery rate. Instead,  all nodes recover simultaneously at some threshold time $T_{max}$, and the epidemic stops. For simplicity, we further assume $T_{max}$ to be fixed, but our methods allow for varying $T_{max}$ for different epidemics.

\subsubsection{Community-Based SI-BD Model (C-SI-BD)}\label{sec:community_model}
Another model that allows for a simpler theoretical analysis is based on the setting from~\citep{sanders2020clustering}. It is assumed that the spreading does not occur over the edges of a graph $G$ but depends solely on the community structure $\A$. As before, the first node of a cascade is chosen uniformly at random. Each activated node can infect all other susceptible nodes independently after an exponentially distributed time. If a susceptible node belongs to the same community, then the infection rate is $\alpha_{in}$; otherwise, it is $\alpha_{out}$, $\alpha_{out}<\alpha_{in}$. Epidemic stops at time  $T_{max}$.

\vspace{10pt}

We want to stress a principal difference between the first two models (SIR and SI-BD) and the C-SI-BD model. C-SI-BD completely ignores the graph structure and takes into account only community assignments. Mathematically, C-SI-BD is equivalent to an SI-BD model on a complete graph, where the infection rate through the intra- and inter- community edges are $\alpha_{in}$ and $\alpha_{out}$, respectively. Then, all nodes within one community are exchangeable, so that at any point of time, the cascade is completely described by the number of infected nodes in each community, and therefore one can hope to recover the community assignments when the number of cascades is large. On the other hand, SIR and SI-BD use explicit information on the structure of the underlying graph. Usually, graph edges are correlated with community assignments, while the infection rate over all edges is the same. As the number of cascades tends to infinity, one can get an asymptotically consistent estimator of
the graph edges, but this might still not be sufficient to recover the communities. A well-known example of this phenomenon is the detectability threshold in the Stochastic Block Model~\cite{decelle2011asymptotic}. In this setup, the quality of recovered communities might not improve beyond certain point, as the number of cascades grows. Such a setup is much more challenging but is realistic.

\section{Algorithms}\label{sec:algorithms}

\subsection{Background on Likelihood Maximization}\label{sec:likelihood}

Recall that $t_i^C$  denotes the activation time for node $i$ in cascade $C$; we often omit the index $C$ when the context is clear. Without loss of generality, for the first node of a cascade, we set $t_i = 0$. Finally, if a node $i$ is not infected during an epidemic, then we set $t_i = T_{max}$. 

Denote $\Delta_{i,j}^C = \Delta_{i,j} = |t_i - t_j|$. 
The log-likelihood $\log L(C)$ of the cascade $C$ for SI-BD with varying infection rates ($i$ infects a susceptible neighbor $j$ after an exponentially distributed time with rate $\alpha_{i,j}$) is given in~\cite{rodriguez2011uncovering}, Equation~(7). For our purposes, it is convenient to write this expression as 
\begin{equation}\label{eq:likelihood_exp}
\log L(C) = - \sum_{i,j: i<j} \alpha_{i,j} \Delta_{i,j}  + \sum_{i: t_i\in(0,T_{max})} \log\sum_{j: t_j < t_i}  \alpha_{j,i}  \,. 
\end{equation}
Here, we substituted a general form of the SI-BD epidemic model to the formula from~\citep{rodriguez2011uncovering} and took the logarithm. 

The log-likelihood for all cascades is $\log L(\C) = \sum_{C \in \C} \log L(C)$. We will next introduce a method based on maximizing $\log L(\C)$ under the cascade-based model.

\subsection{\textsc{ClustOpt} }
\label{sec:clustopt}

Consider the C-SI-BD model described in Section~\ref{sec:community_model}. Denote by $a(i)$ the community assignment for a node $i$. Recall that C-SI-BD is equivalent to the SI-BD model used in~\cite{rodriguez2011uncovering} when the latter is applied to a complete graph with $\alpha_{i,j} = \alpha_{in}$ if $a(i) = a(j)$ and $\alpha_{i,j} = \alpha_{out}$ otherwise. The log-likelihood in Equation~\eqref{eq:likelihood_exp} becomes
\begin{multline}\label{eq:likelihood_communities}
\log L(C,\A) = - (\alpha_{in} - \alpha_{out}) \sum_{\substack{i,j: i<j,\\
a(i) = a(j)}}  \Delta_{i,j} 
- \alpha_{out} \sum_{i,j: i<j}  \Delta_{i,j}
\\
 + \sum_{i:t_i\in(0,T_{max})} \log \left( (\alpha_{in} - \alpha_{out}) |\{j: t_j<t_i, a(j)=a(i)\}| + \alpha_{out} |\{j: t_j<t_i \}| \right)\,.
\end{multline}

Note that in practice, $T_{max}$ is not known, so in the experiments, for each cascade, we set $T_{max}$ to be the time of the last infection plus the average time difference between successive infections.

\vspace{5pt}

We propose the following algorithm
to maximize Equation~\eqref{eq:likelihood_communities}. 
\begin{algorithm}
\caption{\textsc{ClustOpt}}\label{alg:ClustOpt}
\begin{enumerate}[noitemsep,nolistsep]
\item Find initial partition $\A_{init}$;
\item Find $\hat\alpha_{in}, \hat\alpha_{out}
=  \argmax_{\alpha_{in},\alpha_{out}}\log L(\C,\A_{init})$;
\item For fixed $\hat\alpha_{in}, \hat\alpha_{out}$, find $\hat\A = \argmax_{\A}\log L(\C,\A)$.
\end{enumerate}
\end{algorithm}

\begin{remark}
\emph{Algorithm 1 was initially
inspired by the idea of alternated maximization. Surprisingly, our preliminary experiments have shown that repeating steps (2) and (3) multiple times does not improve the performance if we start from a good initial partition $\A_{init}$ in step (1). A possible explanation is that although the maximization in Equation~\eqref{eq:likelihood_communities} does depend on $\alpha_{in}$ and $\alpha_{out}$, it is determined to a great extent by $\Delta_{ij}$. Therefore, we do not need $\hat\alpha_{in}$ and $\hat\alpha_{out}$ to be precise; we just need them to be good enough to find a high-quality $\hat \A$ in step (3). This is achieved when we start with a reasonable initial partition. That is why having a good initial partition is important for stable results.}
\end{remark}

Let us now discuss how the steps (1)--(3) are implemented.

\paragraph{Step (1):} 
The algorithm is initialized with some reasonable initial partition $\A_{init}$.
In the current article, we propose to start from \textsc{Clique(0)} explained in Section~\ref{sec:surrogate}.

\paragraph{Step (2):}
Without loss of generality, we set $\alpha_{in} = (\delta + 1) \alpha_{out}$.
Then, Equation~\eqref{eq:likelihood_communities} becomes
\begin{multline}\label{eq:likelihood_communities-delta}
\log L(C,\A) = - \alpha_{out} \, \delta \sum_{\substack{i,j: i<j,\\ a(i) = a(j)}}  \Delta_{i,j} 
- \alpha_{out} \sum_{i,j: i<j}  \Delta_{i,j} \\
+ \sum_{i: t_i\in(0,T_{max})} \log \left( \delta |\{j: t_j<t_i, a(j)=a(i)\}| + |\{j: t_j<t_i\}| \right)
 + (|C|-1)\log\alpha_{out} \,. 
\end{multline}
Using this, we can find optimal $\alpha_{out}$ in terms of $\delta$ and~$\A$:
\begin{equation}
\label{eq:alpha_hat}
\hat \alpha_{out} = \frac{\sum_{C \in \C} (|C|-1)}{\sum_{C\in \C}\left(\delta \sum_{i,j:a(i) = a(j)}  \Delta_{i,j}^C 
+ \sum_{i,j}  \Delta_{i,j}^C
\right)}.
\end{equation}
Now, we can resort to a numerical solution: For each $\delta$, we set $\alpha_{out}$ as in Equation~\eqref{eq:alpha_hat} and find such $\delta$ that maximizes Equation~\eqref{eq:likelihood_communities-delta}.

\paragraph{Step (3):}
We follow~\cite{prokhorenkova2018community} and adapt the Louvain algorithm for the likelihood given in Equation~\eqref{eq:likelihood_communities} by computing the gain in $\log L(\C,\A)$ obtained by moving a node $v$ from one community to another. Because of computational complexity, we consider only moving single nodes from one community to another and do not attempt to move groups of nodes or merge communities. More formally, we do the following. At the beginning, all vertices are grouped according to $\A_{init}$. Then, we iterate through all vertices: For each vertex $i$, we compute the gain in $\log L(\C,\A)$ coming from putting $i$ to the community of its neighbor and put $i$ to the community with the largest gain, as long as it is positive. We stop the process when we cannot improve $\log L(\C,\A)$ by such local moves.

\vspace{5pt}

\begin{remark} \emph{In the preliminary version of this article~\citep{prokhorenkova2019learning}, a more involved  algorithm called GraphOpt was presented. GraphOpt maximizes the likelihood under the SI-BD model rather than the C-SI-BD model. It uses the expectation-maximization framework assuming that the graph structure is a latent variable. This is a significantly more complicated task because the likelihood must be computed over all possible realizations of the hidden graph. While GraphOpt shows excellent results in some cases, its performance is unstable, and its time complexity is unacceptably large.}
\end{remark}

\subsection{Clustering of Surrogate Graphs}\label{sec:surrogate}

This section presents simple yet effective methods that construct a surrogate graph $\hat G$ and then cluster this graph by some community detection algorithm. We will mostly use the Louvain algorithm~\cite{blondel2008fast}. This method is appealing for several reasons: It is efficient, gives stable results, and does not require the number of clusters as an input (in contrast to, e.g., the spectral clustering).\footnote{We analyze the effect of a community detection algorithm on the performance of surrogate-graph-based methods by comparing the results obtained with Louvain to those obtained using another clustering algorithm, Infomap~\cite{rosvall2008maps} (see Sections~\ref{sec:infomap}, \ref{sec:infomap-exp}).}

It is crucial that $\hat G$ does {\it not} need to be similar to $G$, it just needs to capture the community structure on an aggregated level. Our experiments show that clustering of $\hat G$ often performs better than first inferring $G$ and then clustering it. 

\subsubsection{\textsc{Path} Algorithm} 

For each $C \in \C$, let $G^* = G^*(C)$ be a path connecting subsequently activated nodes. Then, we obtain $\hat G$ by aggregating $G^*(C)$ over all cascades. This leads to the following algorithm.

\begin{algorithm}
\caption{\textsc{Path}}
\label{alg:Path}
\begin{enumerate}[noitemsep,nolistsep]
\item Construct weighted graph $\hat G$, where the weight of each \\ edge $e$ in $\hat G$ is the number of $G^*(C)$ including $e$;\footnotemark
\item Find clusters in $\hat G$ using the Louvain algorithm.
\end{enumerate}
\end{algorithm}
\footnotetext{Weighting edges is important, since for large $|\C|$ unweighted union of edges can be a complete graph, with no meaningful partition.}

While this algorithm cannot be formally justified, it has the following motivation.
Assume that $\C$ is generated  by the SI-BD model. Then, in Equation~\eqref{eq:likelihood_exp}, we set $\alpha_{i,j} = \alpha$ if $i$ and $j$ are connected in $G$ and $\alpha_{i,j} = 0$ otherwise. So, we obtain
\begin{multline}\label{eq:cascade_loglike}
\log \P(C|G) = - \alpha \sum_{(i,j) \in E(G)} \Delta_{i,j} + (|C|-1)\log\alpha + \sum_{i: t_i\in(0,T_{max})} \log |\{j: t_j < t_i, (i,j) \in E(G) \}| 
 \,. 
\end{multline}
Now, for $C \in \C$ let us consider a graph $G'$ with exactly $|C| - 1$ edges that maximizes the probability $\P(C|G')$ in Equation~\eqref{eq:cascade_loglike}. It is easy to check that $G'$ is merely a path connecting subsequently activated nodes, i.e., $G' = G^*$.

\subsubsection{\textsc{Clique}}

Another approach is to include {\it all} possible edges that could participate in the cascade, weighing them by the proxy of their likelihood. Then, each cascade $C$ results in a weighted clique of size $|C|$. The weights can be chosen, for example, as follows. For a cascade $C$ and two nodes $i,j$, let us consider the  probability $P^C(i,j) = \P(\mbox{$j$ was infected from $i$} | C)$. If $t_i>t_j$, then, obviously, $P^C(i,j) = 0$. If $t_i<t_j$, then, as in \citep{gomez2010inferring}, we assume that $P^C(i,j)$ decreases exponentially with $\Delta_{i,j}$; namely, $P^C(i,j) = c_j \, e^{-a\Delta_{i,j}},$ where $c_j$ is a constant depending on $j$.  Since $j$ was infected from exactly one previous node, we must have $\sum_{i: t_i<t_j}P^C(i,j)=1.$ Therefore, 
\begin{equation}\label{eq:clique_prob}
P^C(i,j) = \frac{e^{-a\Delta_{i,j}}}{\sum_{l:t_l<t_j}\,e^{-a\Delta_{l,j}}}\,.
\end{equation}
Parameter $a$ essentially balances between paths and cliques: If $a$ is large, we mostly take into account subsequent nodes with small $\Delta_{i,j}$; for small $a$ all pairs of nodes participated in $C$ are important.

\begin{remark}
\emph{Note that we directly model $P^C(i,j)$ without making assumptions about the distribution of $\Delta_{i,j}'s$. Although $P^C(i,j)$ in Equation~\eqref{eq:clique_prob} is proportional to the exponential density function $ae^{-ax}$ for $x=\Delta_{i,j}$, recall that if we assume that infection times are exponentially distributed, then a node is equally likely to get infected from any of the currently infected nodes due to the memory-less property. Furthermore, similarly to \textsc{Path}, this algorithm cannot be formally justified as optimization under a particular generative model. Indeed, if we assume a latent graph structure, then the expressions for the a-posteriori probabilities, given $C$, that an edge exists, or that a node has been infected through this edge, are much more involved than the simple expression \eqref{eq:clique_prob}; see, e.g., the GraphOpt algorithm in \cite{prokhorenkova2019learning}.} 
\end{remark}

\textsc{Clique} constructs $\hat G$ as the weighted graph with weight of $\{i,j\}$ given by $\sum_{C\in \C} (P^C(i,j)+P^C(j,i))$, which, under our assumptions, is the expected number of times infections passed between $i$ and $j$.  To make \textsc{Clique} insensitive to the speed of epidemics, we take $a = \frac{1}{\Delta}$, where $\Delta$ is the average time between infection times (we average over all pairs of infected nodes belonging to the same cascade). We also consider \textsc{Clique(0)} with $a = 0$, which is a natural choice for the SI-BD model because it mimics the memory-less property of exponential infection times. We noticed that \textsc{Clique}  is not too sensitive to varying $a$ in a reasonable interval, while for too large $a$ \textsc{Clique} becomes very similar to \textsc{Path}.

\begin{algorithm}
\caption{\textsc{Clique}}\label{alg:Clique}
\begin{enumerate}[noitemsep,nolistsep]
\item Construct weighted graph $\hat G$ with weight of $\{i,j\}$ given by $\sum_{C\in \C} (P^C(i,j)+P^C(j,i))$;
\item Find clusters in $\hat G$ using the Louvain algorithm.
\end{enumerate}
\end{algorithm}

\subsubsection{\textsc{CosineSim}}\label{sec:cosinesim}

This algorithm is motivated by~\citet{ramezani2018community}, where the authors use cosine similarity to get initial similarities of nodes. Then, instead of applying the procedure from~\cite{ramezani2018community}, we apply the Louvain algorithm to the obtained graph. 

\begin{algorithm}
\caption{\textsc{CosineSim}}\label{alg:CosineSim}
\begin{enumerate}[noitemsep,nolistsep]
\item Compute $V_i = \left(\1_{\{v_i \in C_1, \ldots, v_i \in C_r\}}\right)_{j=1}^r$;
\item Construct $\hat G$ as the weighted graph with weight of $\{i,j\}$ given by $\frac{\langle V_i, V_j \rangle}{|V_i| |V_j|}$;
\item Find clusters in $\hat G$ using the Louvain algorithm.
\end{enumerate}
\end{algorithm}

\subsection{Baselines}
\label{sec:baselines}

\textsc{MultiTree} uses the algorithm from~\cite{rodriguez2012submodular} to find an inferred graph $\hat G$, which is then clustered by the Louvain algorithm~\citep{blondel2008fast}.

We also propose \textsc{Oracle} as a superior benchmark for all possible network inference algorithms. Specifically, we construct a graph $\hat G$ consisting of all edges that participated in cascades (assuming these edges are known, hence, the name \textsc{Oracle}). We cluster the obtained graph $\hat G$ by the Louvain algorithm. Let us stress that we do not consider \textsc{Oracle} as our competitor as it uses information that is not available to other algorithms.

Finally, we used algorithms proposed for solving the same problem as in our work: \textsc{C-IC} and \textsc{C-Rate}~\citep{barbieri2017efficient}, as well as \textsc{R-CoDi} and \textsc{D-CoDi}~\citep{ramezani2018community}. We use the publicly available implementations provided by the authors. 

\begin{figure}[htb]
    \centering
    \includegraphics[width=0.8\textwidth]{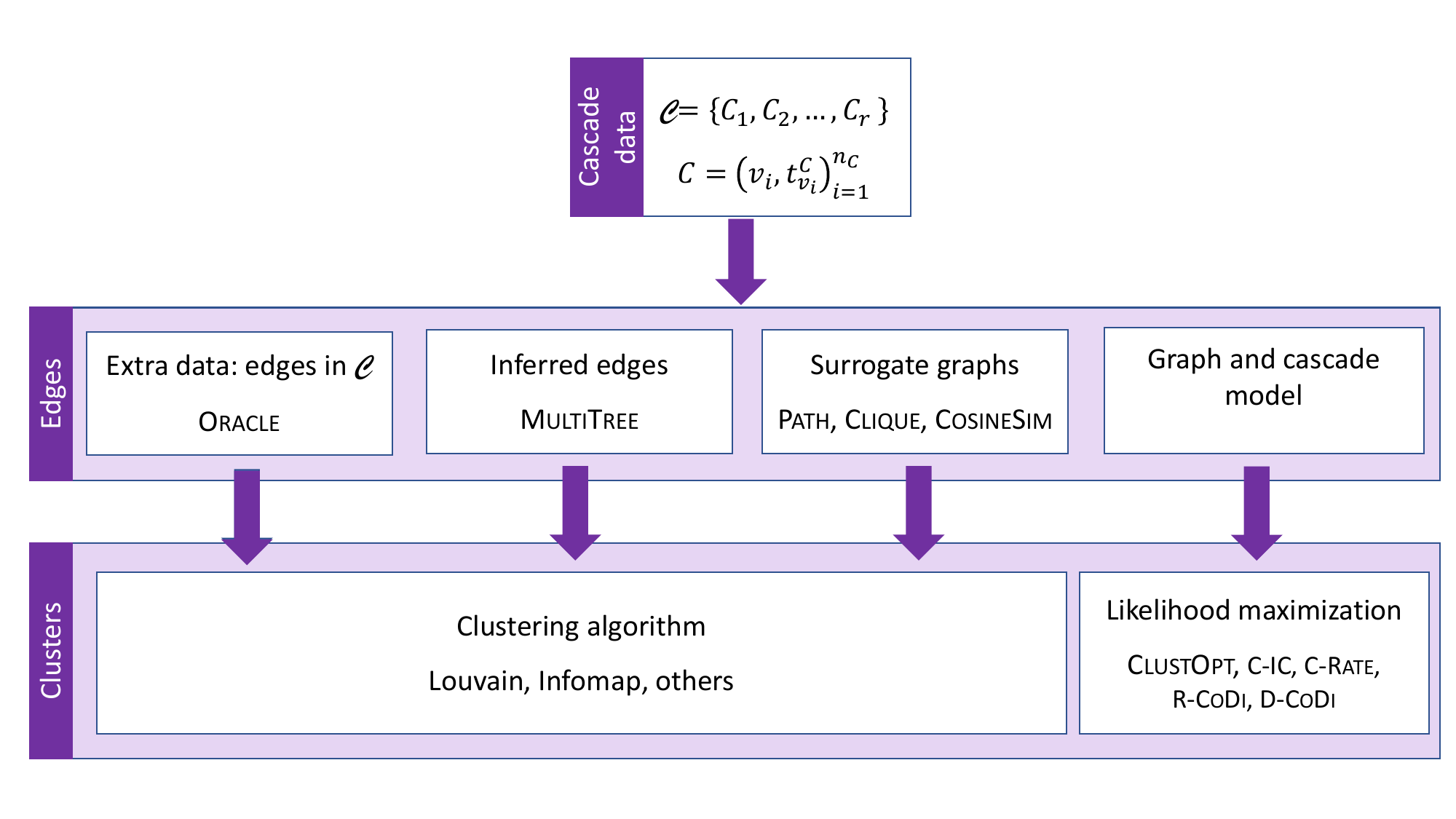}
    \caption{Overview of the algorithms. The proposed methods are \textsc{Path, Clique, ClustOpt, and CosineSim}. Baselines are: \textsc{C-IC} and \textsc{C-Rate}~\citep{barbieri2017efficient},  \textsc{R-CoDi} and \textsc{D-CoDi}~\citep{ramezani2018community}, and \textsc{MultiTree}~\cite{rodriguez2012submodular}. \textsc{Oracle} is a superior benchmark for all network inference algorithms.}
    \label{fig:algorithms-overview}
\end{figure}

A summary of the proposed algorithms and the baselines is schematically represented in  Figure~\ref{fig:algorithms-overview}. Stage ``Edges'' refers to constructions that connect the cascade data to a (possibly surrogate) network. Stage ``Clusters'' refers to obtaining communities.

\section{Experimental Setup}\label{sec:setup}

\subsection{Datasets}\label{sec:networks}\label{sec:datasets}

For a fair comparison of all algorithms, we performed a series of experiments using synthetic and real-world datasets.

\begin{table}[t]
\vspace{5pt}
\caption{Twitter Dataset}
  \label{tab:twitter_dataset}
  \centering
  \begin{tabular}{lcccc}
Dataset&$n$&$k$&$r$&$\mathrm{avg}(n_C)$ \\
    \hline
    Twitter
    	& 18,470 & 2 & 29,998 & 3\\
  \hline
\end{tabular}
\end{table}

The main dataset for our experiments is the collection of Twitter posts obtained from September till November 2010~\citep{conover2011political}. 
Importantly, this dataset combines both community assignments and cascades.
For community assignments, Twitter users are manually classified by political alignment.
The original dataset contains the following information about retweets: id of a user that started the tweet, id of a user that retweeted, the timestamp, the list of hashtags, and the number of hyperlinks included in the tweet. We processed the dataset to extract explicit cascades: Although no cascade ids are present, we assumed that a cascade can be uniquely identified by the list of hashtags, source user id, and the number of hyperlinks. Additionally, the timestamp of the first tweet is not known, so we assumed that the time difference between the original tweet and the first retweet is the same as between the first and the second retweets. This way, we obtained a dataset with about 18.5K nodes and 30K cascades. (We did not perform any further filtering.) The dataset statistics are summarized in Table~\ref{tab:twitter_dataset}.

The second group of datasets contains real-world networks with ground truth community assignments (see Table~\ref{tab:datasets}). For these datasets, we use generated epidemics (see Section~\ref{sec:cascades}) and present the results aggregated over the datasets (see Section~\ref{sec:average}). We include many datasets of different nature and sizes because we want to show that our conclusions generalize to different scenarios.

Finally, we also add a synthetic dataset based on the LFR model~\citep{lancichinetti2008benchmark}, which is a standard synthetic benchmark for networks with communities. This model generates graphs with power-law distributions of node degrees and cluster sizes and includes the mixing parameter $\mu$, which is the fraction of inter-cluster edges. We generated a graph on 10K nodes with the power-law exponent of the degree distribution 2.5, the power-law exponent of the cluster size distribution 1.5, mixing parameter $\mu = 0.1$, average degree 5, maximum degree 100, and community sizes between 100 and 600. (We obtained five clusters in total.)

\begin{minipage}{0.48\textwidth} 
\vspace{5pt}
\captionof{table}{Real-World Datasets}
  \label{tab:datasets}
  \centering
  \begin{tabular}{lccc}
Dataset&$n$&$m$&$k$ \\
    \hline
    Karate club~\citep{zachary1977information} 
    	& 34 & 78 & 2 \\
    Dolphins~\citep{lusseau2003bottlenose} 
    	& 62 & 159 & 2 \\
	Football~\citep{newman2004finding} 
    	& 115 & 613 & 11 \\
        	Political books~\citep{newman2006modularity} & 105 & 441 & 3 \\
    email-Eu-core~\citep{leskovec2007graph} & 986 & 16,064 & 42 \\
    Newsgroup~\citep{yen2007graph} & 999 & 10,194 & 5 \\
	Political blogs~\citep{adamic2005political} 
    	& 1,224 & 16,715 & 2  \\
    Cora~\citep{mccallum2000automating} & 2,708 & 5,279 & 7 \\
    CiteSeer~\citep{giles1998citeseer} & 3,264 & 4,536 & 6 \\
  \hline
\end{tabular}
\end{minipage}
\begin{minipage}{0.48\textwidth} 
\vspace{5pt}
\captionof{table}{Parameters Used for Synthetic Epidemics}
\label{tab:parameters}
  \centering
  \begin{tabular}{lcccc}
Dataset&$\alpha$&$\alpha_{in}$&$\alpha_{out}$&Lomax \\
    \hline
    Karate 
    	& 0.15 & 0.09 & 0.009 & 12 \\
    Dolphins 
    	& 0.14 & 0.047 & 0.0047 & 14 \\
	Football 
    	& 0.07 & 0.083 & 0.0083 & 32 \\
        	Pol-books & 0.08 & 0.036  & 0.0036 & 26 \\
    Eu-core & 0.016 & 0.012 & 0.0012 & 230 \\
    Newsgroup & 0.032 & 0.0058 & 0.00058 & 105 \\
	Pol-blogs 
    	& 0.017 & 0.0024 & 0.00024 & 216  \\
    Cora & 0.14 & 0.0024 & 0.00024 & 23 \\
    CiteSeer & 0.21 & 0.0019 & 0.00019 & 16 \\
  \hline
\end{tabular}
\end{minipage}
\vspace{8pt}

\subsection{Synthetic Cascades}\label{sec:cascades}

For the LFR graph and the datasets described in Table~\ref{tab:datasets}, we generated the cascades according to all models discussed in Section~\ref{sec:cascade_models}. Importantly, there are both edge-based and community-based models.

\begin{figure}
\begin{center}
\includegraphics[width=.49\textwidth]{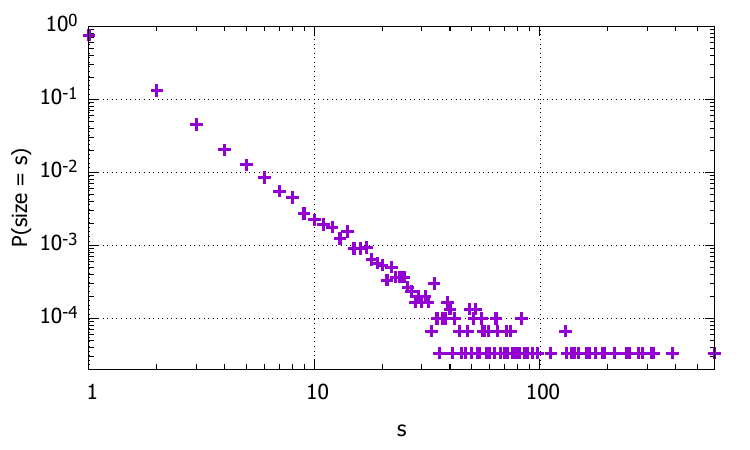}
\includegraphics[width=.49\textwidth]{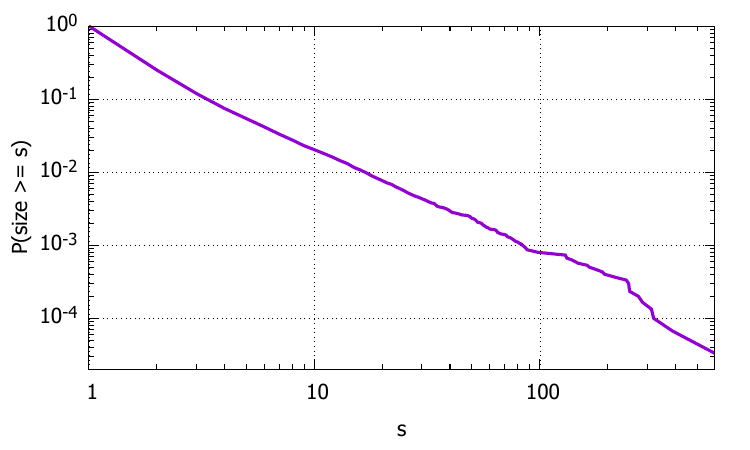}
\end{center}
\caption{Distribution of cascade sizes, Twitter dataset.}
\label{fig:twitter_dist}
\end{figure}

\begin{figure}
\begin{center}
\includegraphics[width=.49\textwidth]{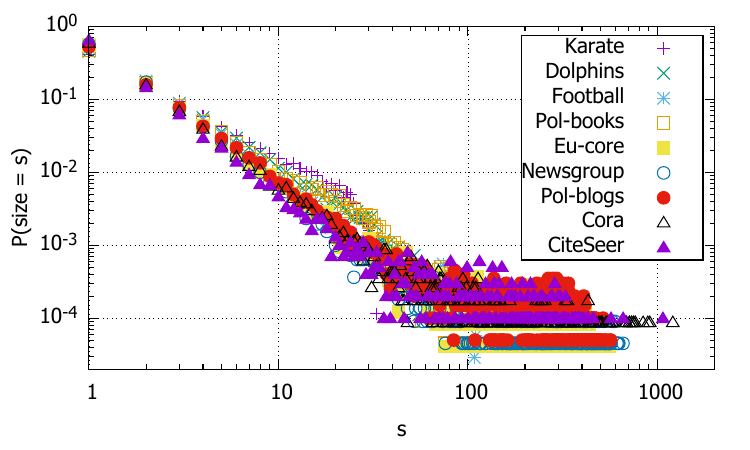}
\includegraphics[width=.49\textwidth]{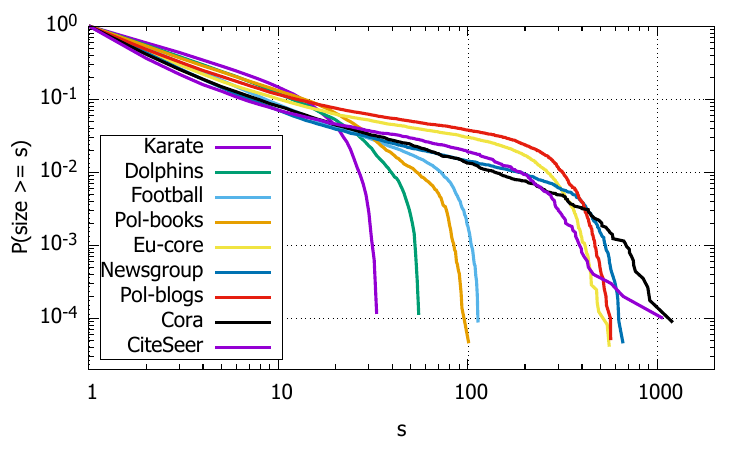}
\end{center}
\caption{Distribution of cascade sizes, SIR model: density (left) and complementary cumulative distribution function (right).}
\label{fig:SIR}
\end{figure}

\begin{figure}
\begin{center}
\includegraphics[width=.49\textwidth]{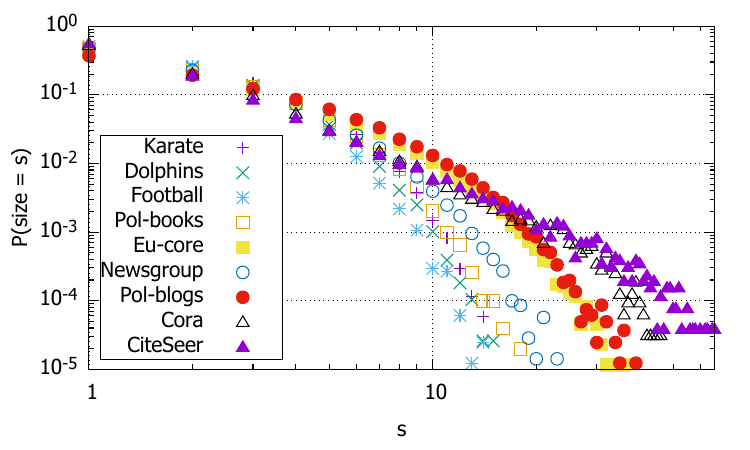}
\includegraphics[width=.49\textwidth]{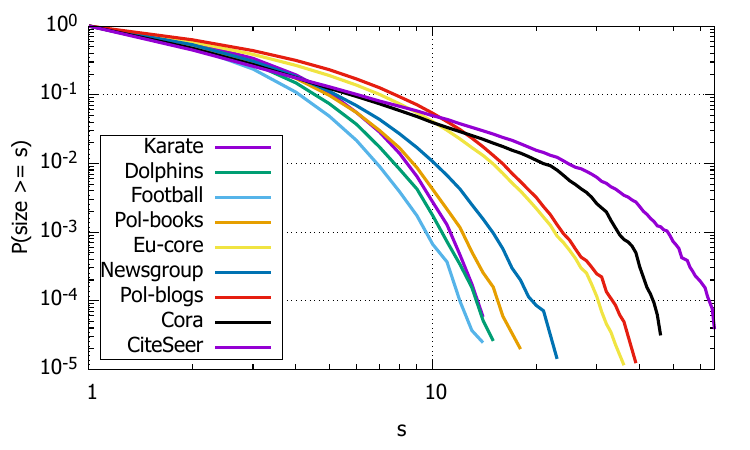}
\end{center}
\caption{Distribution of cascade sizes, SI-BD model: density (left) and complementary cumulative distribution function (right).}
\label{fig:SI-BD}
\end{figure}

\begin{figure}
\begin{center}
\includegraphics[width=.49\textwidth]{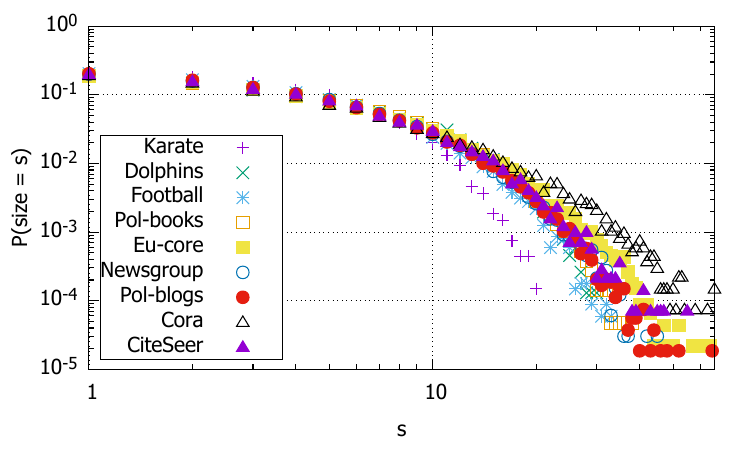}
\includegraphics[width=.49\textwidth]{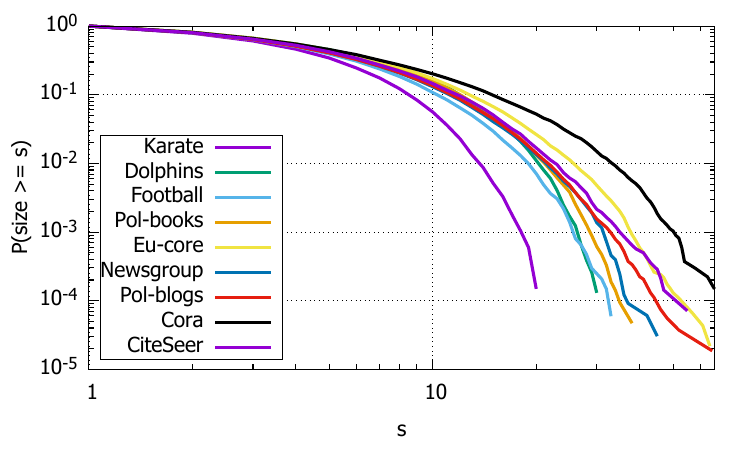}
\end{center}
\caption{Distribution of cascade sizes, C-SI-BD model: density (left) and complementary cumulative distribution function (right).}
\label{fig:C-SI-BD}
\end{figure}

Without loss of generality, we set $\beta = 1$ for the SIR model and $T_{max} = 1$ for SI-BD and C-SI-BD. It remains to choose the parameters $\alpha$, $\alpha_{in}$, $\alpha_{out}$, and the parameter of the Lomax distribution used by SIR. 
We noticed that these parameters have to be carefully chosen to get an informative set of cascades. Otherwise, we may either get cascades consisting of single nodes or cascades consisting of almost all nodes. The following heuristics work well enough:
For SIR and SI-BD, we choose parameters so that the average size of a cascade is 2; for C-SI-BD, we take $\alpha_{in} = 10 \,\alpha_{out}$ and choose $\alpha_{out}$ such that the number of cascades consisting of one node is about 20\%. 
In Table~\ref{tab:parameters}, we list the parameters we used for all datasets.
We plotted the obtained distributions of cascade sizes (see Figures~\ref{fig:SIR}--\ref{fig:C-SI-BD}) and checked that they are similar to those observed in the literature for real data~\citep{bakshy2011identifying,galuba2010outtweeting,gomez2013modeling,kwak2010twitter,lerman2012social} and for our Twitter dataset (see Figure~\ref{fig:twitter_dist}).

Finally, we note that  single-node cascades were removed from the generated epidemics.

\subsection{Performance Measures}\label{sec:metrics}

In this section, we discuss evaluating the quality of the obtained communities compared to the ground truth. There are many similarity indices used in the literature for the comparison of partitions. We refer to a recent work by~\citet{gosgens2019systematic} for a systematic study and comparison of popular indices. A critical drawback of many existing similarity measures is that they are biased with respect to cluster sizes, i.e., they prefer too small or too large clusters. For example, the Normalized Mutual Information (NMI)~\citep{bagrow2008evaluating,fortunato2010community} is known to prefer smaller communities. 

Our particular task of evaluating cascade-based community detection has some additional difficulties.
Indeed, let $V_0\subset V$ denote the set of all nodes that participated in cascades. In many realistic scenarios, $V_0$ contains only a small fraction of nodes in $V$. As a result, we obtain a partition of $V_0$, which has to be compared to the ground truth partition of $V$.
There are two possible approaches to make the comparison: (1) restrict the ground truth partition only to $V_0$ and (2) extend the obtained partition from $V_0$ to $V$, e.g., by assigning all nodes in $V\setminus V_0$ to one cluster or individual clusters. Note that the second option has the advantage of being more interpretable (it shows how close we are to the whole ground truth partition). However, at the same time, it is vulnerable to biases: If a similarity index favors too large or too small clusters, then the assignments for the unknown set $V\setminus V_0$ may significantly affect the obtained value. 

To demonstrate the importance of choosing a proper similarity index, we perform a set of experiments comparing different choices:
\begin{itemize}
    \item \textbf{Pearson correlation coefficient} between the incidence vectors is an unbiased similarity index advised by~\citet{gosgens2019systematic}.\footnote{Note that there are other good indices advised in~\citep{gosgens2019systematic}, e.g., the S\&S index. We use the correlation coefficient as it is more widely used.} \emph{Unbiased} means that it does not give a preference to either large or small communities. {\it Pearson-sub} is computed only using the pairs from $V_0$: We measure the Pearson correlation coefficient between the vectors of length $\binom{|V_0|}{2}$ consisting of 0s and 1s, where 1 means that two nodes belong to the same community and 0 means that they belong to different ones. {\it Pearson-all} is computed on the whole dataset: We measure the correlation between two vectors of length $\binom{|V|}{2}$. For the ground truth partition, we know all entries of the vector. In contrast, for our partition, we say that if $u \in V_0$ and $v \in V\setminus V_0$, then the corresponding entry is 0 (different clusters), and if $u,v \in V\setminus V_0$, then the corresponding entry is 0.5 since we do not have any information about nodes in $V \setminus V_0$.
    \item \textbf{NMI} (normalized mutual information)~\citep{bagrow2008evaluating,fortunato2010community} is a widely used similarity measure, which is known to prefer smaller communities. {\it NMI-sub} is computed only on $V_0$, to compute {\it NMI-all} we assign all nodes in  $V\backslash V_0$ to one cluster labeled ``unknown''.
    \item \textbf{Jaccard index} is known to be biased towards larger clusters. {\it Jaccard-sub} is computed on $V_0$, and to compute {\it Jaccard-all}, we say that all unknown nodes belong to \textit{different} communities (since marking them as one community would unreasonably increase the index).
    \item \textbf{F-measure}~\citep{gregory2008fast} is a harmonic average between precision and recall, where precision and recall are defined in terms of node pairs. When we compute \emph{F-measure-all}, we say that all unseen nodes belong to different communities. 
\end{itemize}

In most experiments, we use Pearson-sub as a primary measure since it is proven to be unbiased~\citep{gosgens2019systematic}. However, for the Twitter dataset, we perform a comparison of all listed measures.

\subsection{Aggregating Results over Datasets}\label{sec:average}

In the experiments, we compare the algorithms using all datasets from Table~\ref{tab:datasets}, so it is essential to be able to aggregate the obtained results so that we can compare the overall performance of algorithms over a range of datasets. Moreover, we also want to show the dependence of the performance on the number of available cascades. However, the same number of cascades (e.g., 1,000) can be sufficient to recover communities on a small dataset like Karate, while it can be far from enough to recover communities on a larger CiteSeer dataset. The following heuristic allows for balancing the data: We define a \emph{relative size} of cascades $S$ as the fraction of the overall number of transmissions of infections to the number of edges in the graph. In other words, $S$ is the average number of transmissions through an edge in the graph. We empirically checked that this is a reasonable rescaling since the saturation of quality happens at similar values of $S$ for all datasets.

For a fixed value of $S$, we average the performance over the datasets: We compute the average value of Pearson-sub and the algorithm's average rank. To compute the latter value, for each dataset, we order all algorithms according to their performance (Pearson-sub) and obtain ranks for each algorithm; then, the obtained ranks are averaged over the datasets.

\subsection{Effect of Clustering Algorithm on Surrogate-Graph-Based Methods}
\label{sec:infomap}
As discussed above, we use the Louvain method to cluster surrogate graphs. It turns out that the Louvain algorithm is well suited for clustering of surrogate graphs, and not all algorithms have this property. To show that, we compare Louvain with another well-known community detection method called Infomap~\cite{rosvall2008maps}, which finds communities by minimizing the description length of random walks in the network. The results of the comparison are discussed in Section~\ref{sec:infomap}.

\section{Experiments}\label{sec:experiments}

\subsection{Twitter Dataset}

\begin{figure}
\begin{center}
\includegraphics[width=.49\textwidth]{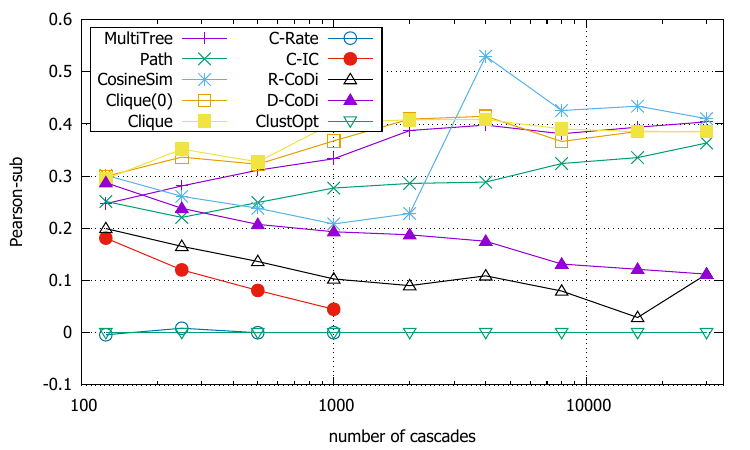}
\includegraphics[width=.49\textwidth]{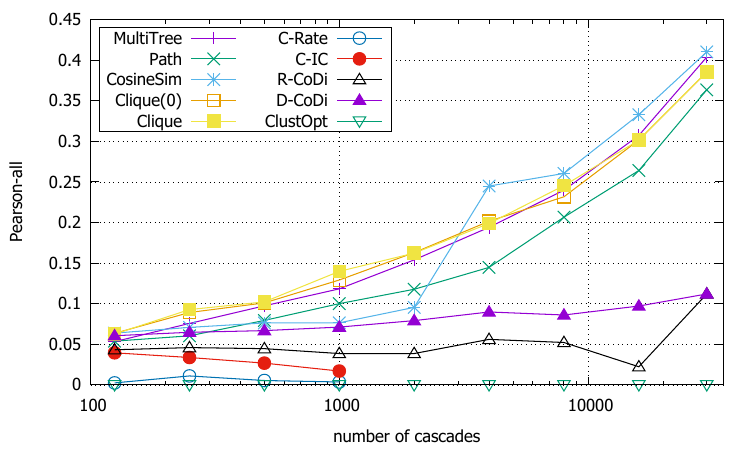}
\includegraphics[width=.49\textwidth]{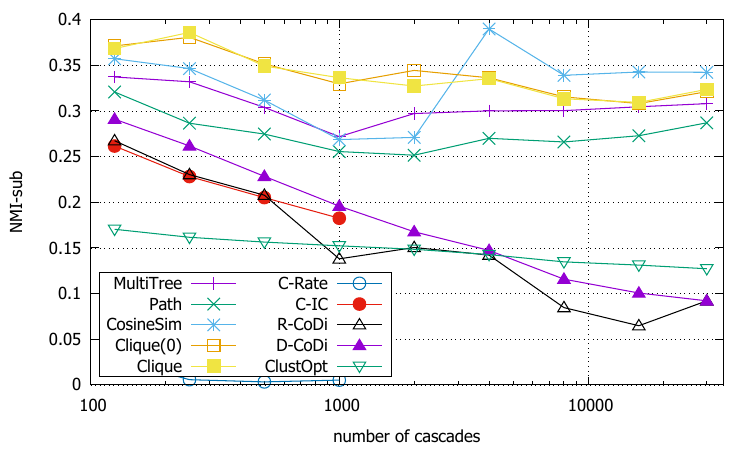}
\includegraphics[width=.49\textwidth]{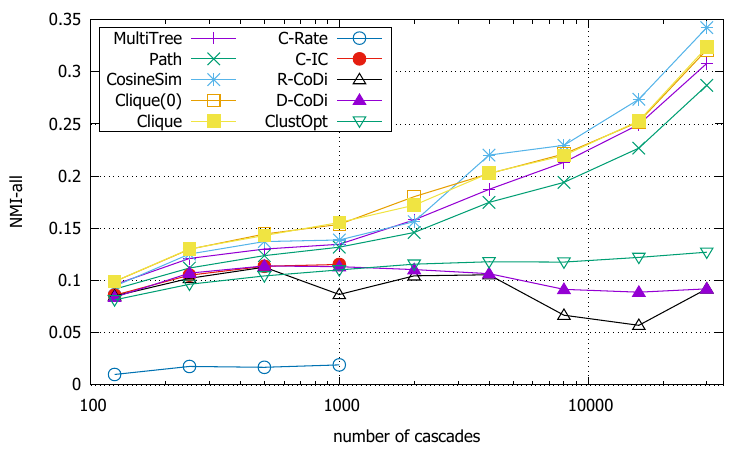}
\includegraphics[width=.49\textwidth]{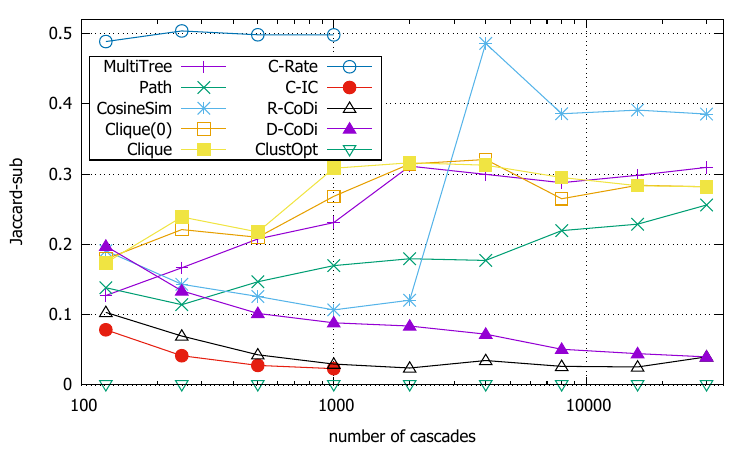}
\includegraphics[width=.49\textwidth]{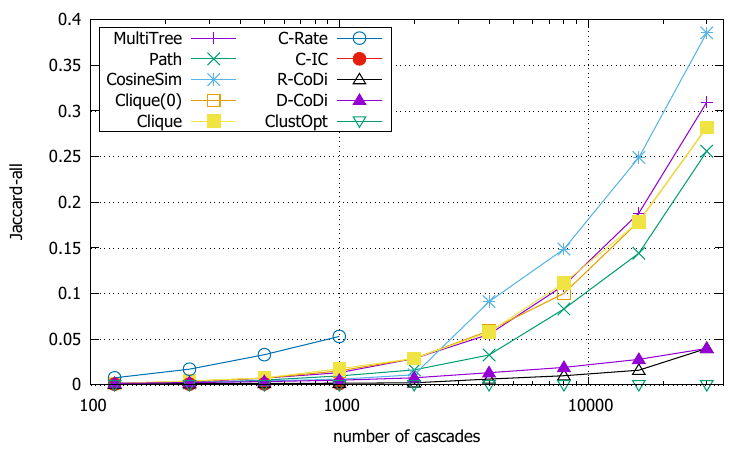}
\includegraphics[width=.49\textwidth]{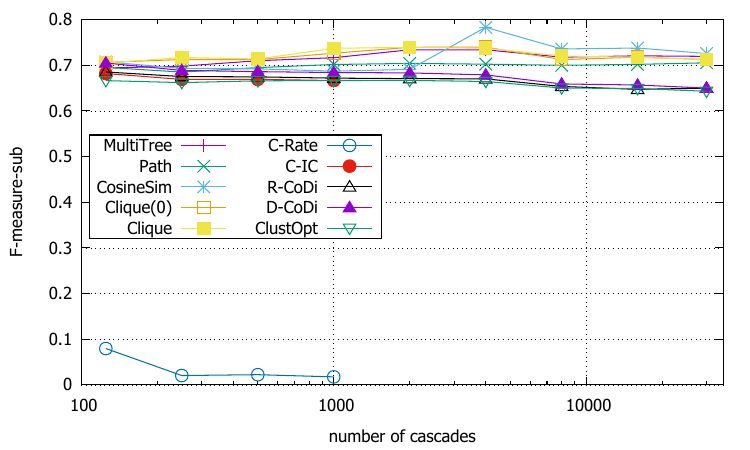}
\includegraphics[width=.49\textwidth]{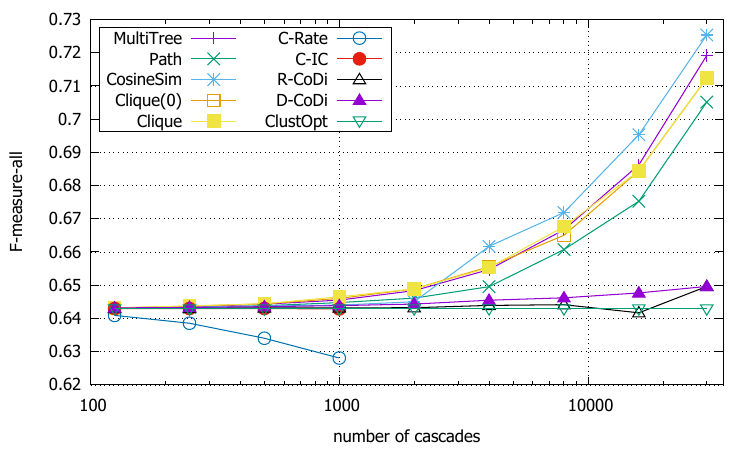}
\end{center}
\caption{Results on the Twitter dataset. The main performance measure for our experiments is Pearson-sub; other measures are presented for illustrative purposes.}
\label{fig:twitter}
\end{figure}

First, we compared all the algorithms on the Twitter dataset, which contains both real cascades and real community assignments; the results are presented in Figure~\ref{fig:twitter}.

Recall that our main performance measure is Pearson-sub (the top-left plot). First, we note that \textsc{Clique} algorithms are stably good for all amounts of cascade data. Further experiments will also confirm this conclusion. Also, \textsc{Clique} and \textsc{Clique(0)} have similar quality. Surprisingly, \textsc{CosineSim} outperforms \textsc{R-CoDi} and \textsc{D-CoDI}, i.e., Louvain algorithm above the similarity graph outperforms a more complicated algorithm presented by~\citet{ramezani2018community}. We guess that there can be two possible reasons for that: (1) \textsc{R-CoDi} and \textsc{D-CoDI} are based on some model assumptions and (2) these algorithms require some parameter tuning, while we used the default parameters suggested by the authors. In contrast, the Louvain algorithm is readily applied without the need for parameter tuning. For \textsc{CosineSim}, we also observed a phase transition when the number of epidemics is 4,000: At this point, \textsc{CosineSim} becomes the best one among all the algorithms. However, this transition is not observed in other experiments.  We also noticed that on this dataset, \textsc{MultiTree} works surprisingly well, especially for large cascade sizes. However, again, this is not supported by our further experiments. Interestingly, the advanced algorithms \textsc{C-IC}, \textsc{C-Rate}, and  \textsc{ClustOpt} have the worst performance. A possible reason is that they are strongly based on model assumptions, which may not hold for real data. Moreover, \textsc{C-IC} and \textsc{C-Rate} could not handle datasets with a large number of cascades. Concerning \textsc{ClustOpt}, this algorithm needs more cascade data to have a good performance, as we show by further experiments.

Now let us demonstrate the problem of proper quality evaluation as discussed in Section~\ref{sec:metrics}. In Figure~\ref{fig:twitter}, in addition to the primary performance measure Pearson-sub, we also show other alternatives. Recall that the measures with the postfix \emph{-all} measure the quality on the same set of nodes for all cascade sizes. For a reasonable measure and a reasonable algorithm, the quality is expected to be monotone with the number of cascades. Note that with 32K cascades, \emph{-all} and \emph{-sub} variants coincide because this number of cascades is sufficient to cover all nodes. By further examining the measures in Figure~\ref{fig:twitter}, we observe huge differences between the results for different measures. These differences can be explained by the fact that some measures are biased, i.e., give a higher preference to either smaller or larger clusters. For example, the \textsc{C-Rate} algorithm is the best according to Jaccard, but the worst according to NMI and F-measure. This inconsistency can be explained by the fact that \textsc{C-Rate} produces too large communities favored by Jaccard, while F-measure and NMI are biased towards smaller clusters.
Similarly, while \textsc{ClustOpt} has zero Pearson correlation with the target clustering, it may outperform other algorithms according to NMI just because it produces many small clusters.
We conclude that it is essential to use unbiased measures. Therefore, we will continue to use the Pearson correlation coefficient in our experiments.

\begin{figure}
\begin{center}
\includegraphics[width=.7\textwidth]{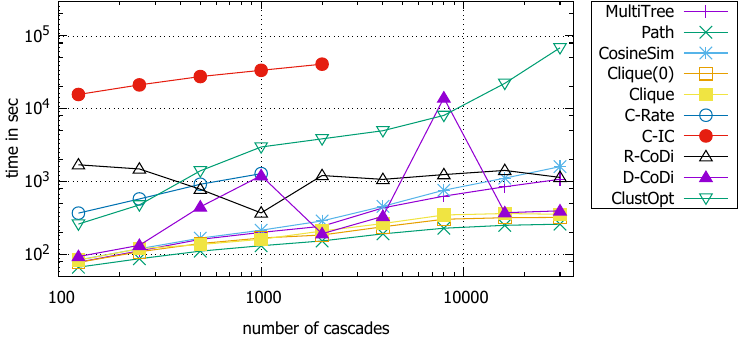}
\end{center}
\caption{Time complexity.}
\label{fig:time}
\end{figure}

We also compared the time complexity of all the algorithms (see Figure~\ref{fig:time}). The algorithms were run on Intel(R) Xeon(R) CPU E5-2650 v2 @ 2.60 GHz.
Note that this result is not an entirely fair comparison of time complexities since the algorithms are implemented via different programming languages with different levels of optimization and parallelization.
Our aim is only to give an intuition about how the complexities scale with the number of cascades. 

\subsection{Other Real-World Datasets}

\begin{figure}[t]
\begin{center}
\includegraphics[width=.49\textwidth]{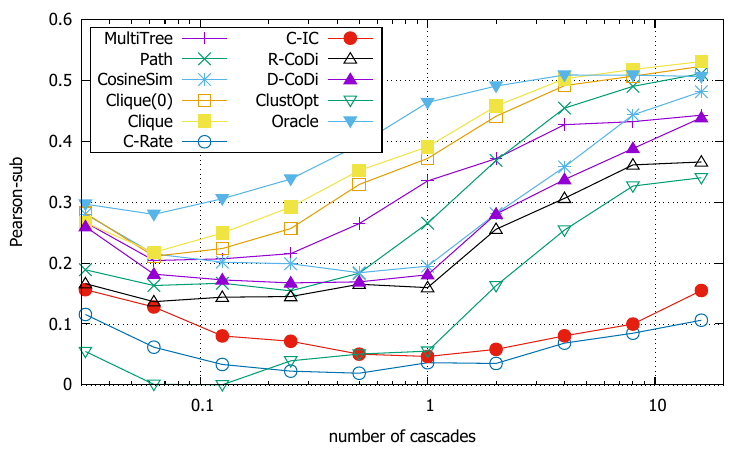}
\includegraphics[width=.49\textwidth]{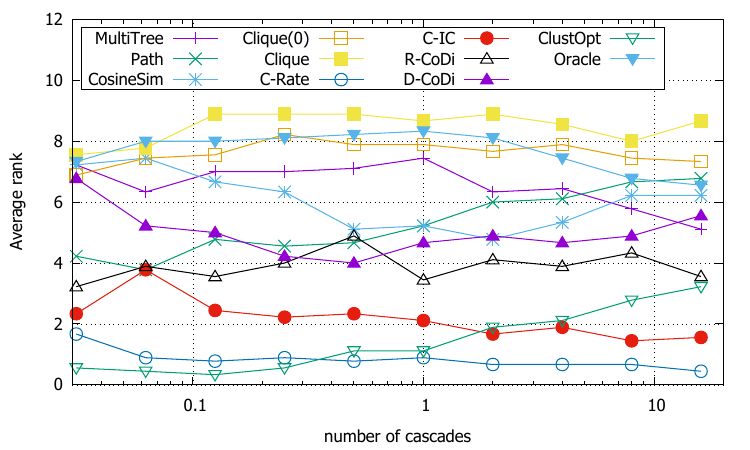}
\end{center}
\caption{Average results on datasets from Table~\ref{tab:datasets}, SIR cascade model.}
\label{fig:real-SIR}
\end{figure}

\begin{figure}[t]
\begin{center}
\includegraphics[width=.49\textwidth]{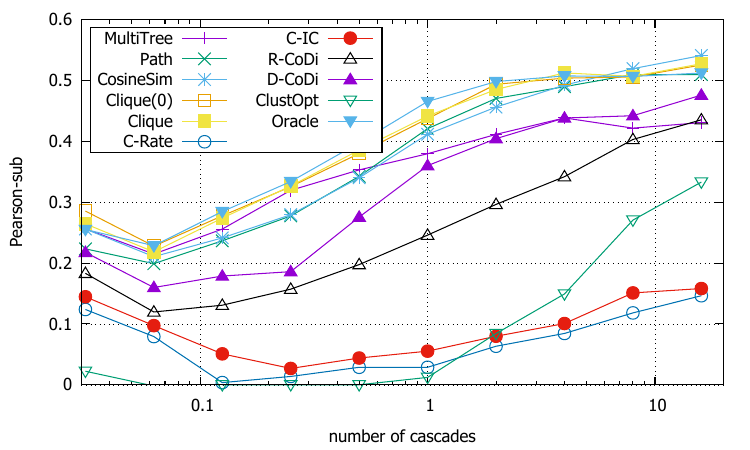}
\includegraphics[width=.49\textwidth]{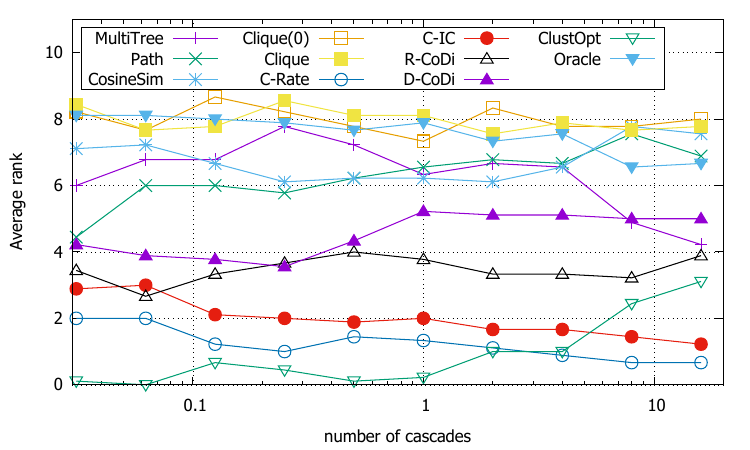}
\end{center}
\caption{Average results on datasets from Table~\ref{tab:datasets}, SI-BD cascade model.}
\label{fig:real-SI-BD}
\end{figure}

\begin{figure}[t]
\begin{center}
\includegraphics[width=.49\textwidth]{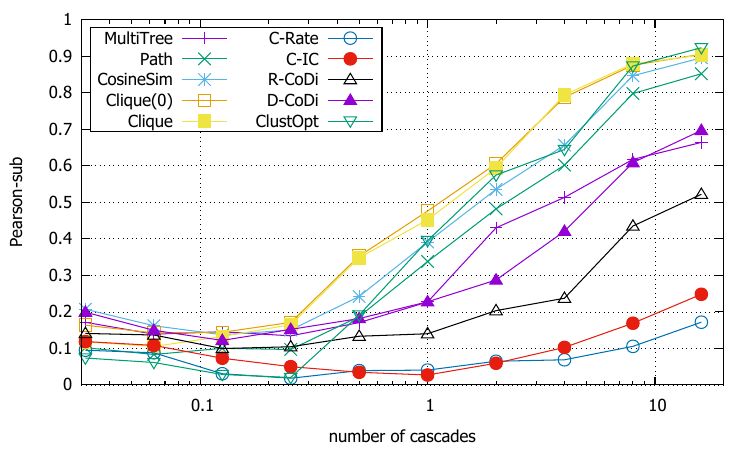}
\includegraphics[width=.49\textwidth]{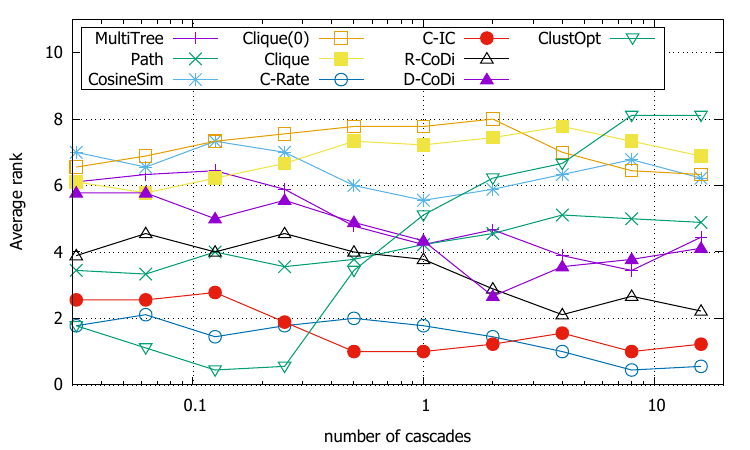}
\end{center}
\caption{Average results on datasets from Table~\ref{tab:datasets}, C-SI-BD cascade model.}
\label{fig:real-C-SI-BD}
\end{figure}

We compared all algorithms on real-world datasets listed in Table~\ref{tab:datasets}. Recall that for these datasets, the epidemics are synthetic and generated according to the models discussed in Section~\ref{sec:cascade_models}. In Figures~\ref{fig:real-SIR}--\ref{fig:real-C-SI-BD}, we plot the average Pearson-sub and the average rank, as explained in Section~\ref{sec:average}.

The results mostly agree with conclusions made on the Twitter dataset. \textsc{Clique} algorithms are the best for all sizes of cascades and all epidemic models. As before, \textsc{CosineSim} outperforms both \textsc{R-CoDi} and \textsc{D-CoDI}. However, there is no phase transition for \textsc{CosineSim} on these datasets, and for all volumes of cascade data, it is outperformed by \textsc{Clique} and \textsc{Clique(0)}. Also, as we discussed above, the performance of \textsc{MultiTree} is not as good as on the Twitter dataset. Speaking about \textsc{ClustOpt}, its performance increases with the number of available cascades. \textsc{ClustOpt} is the best algorithm for the C-SI-BD cascade model when the number of cascades is sufficiently large, which confirms that this algorithm heavily relies on model assumptions and should be replaced by more robust algorithms when there is no evidence that the model assumptions hold.

Finally, recall that \textsc{Oracle} was initially considered as an upper bound for all possible network inference algorithms since it uses the exact information about who infected whom. If the number of cascades is large enough, then \textsc{Oracle} essentially clusters the original graph. Interestingly, in some cases, \textsc{Oracle} is beaten by the surrogate-graph-based algorithms, especially for a large number of cascades, although the surrogate graphs are clustered with the same Louvain algorithm. (The difference is especially apparent on the plots showing the average rank.) One can conclude that a clever weighting of edges can improve the Louvain algorithm. In other words, although Louvain is not an ideal clustering method, the effect of the errors made by Louvain is reduced by the weighting of edges provided by the surrogate-graph-based algorithms.


\subsection{Synthetic Graph}

\begin{figure}
\begin{center}
\includegraphics[width=.49\textwidth]{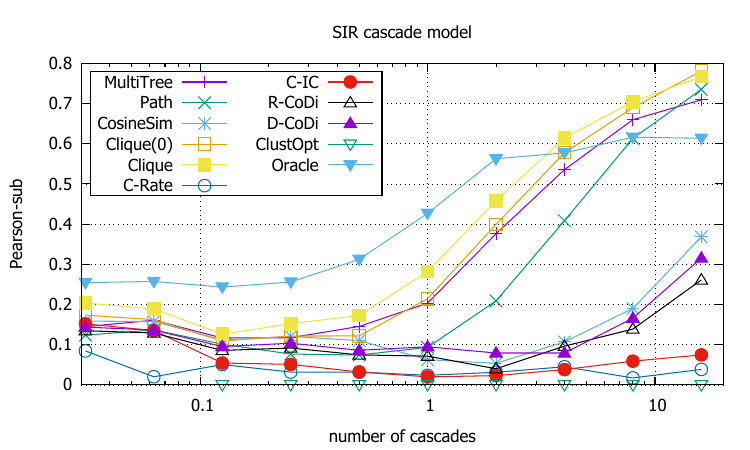}
\includegraphics[width=.49\textwidth]{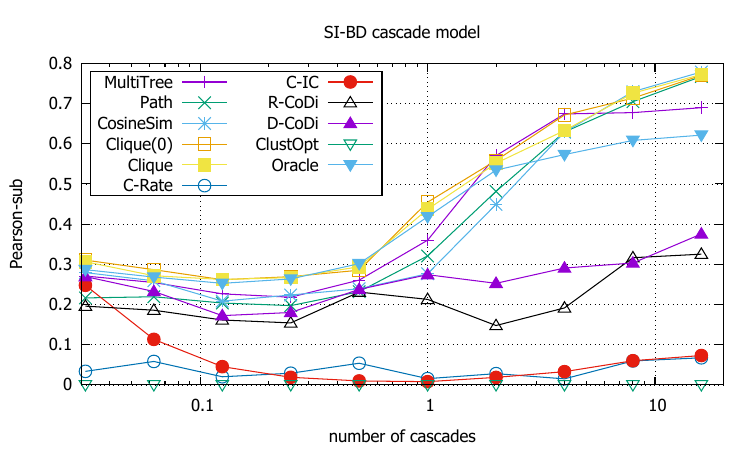}
\includegraphics[width=.49\textwidth]{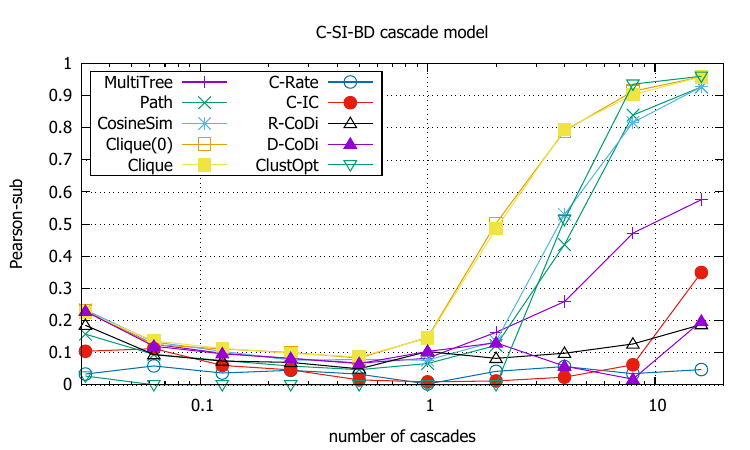}
\end{center}
\caption{Results on a synthetic graph.}
\label{fig:LFR}
\end{figure}

The results on the synthetic LFR model are shown in Figure~\ref{fig:LFR}, which confirms the conclusions made previously on the real-world data. Here it is clearly seen that \textsc{Clique} has good and stable performance and that \textsc{Oracle} can be beaten by others if the number of cascades is large.

\subsection{Clustering of Surrogate Graphs with Infomap}
\label{sec:infomap-exp}

\begin{figure}
\begin{center}
\includegraphics[width=.49\textwidth]{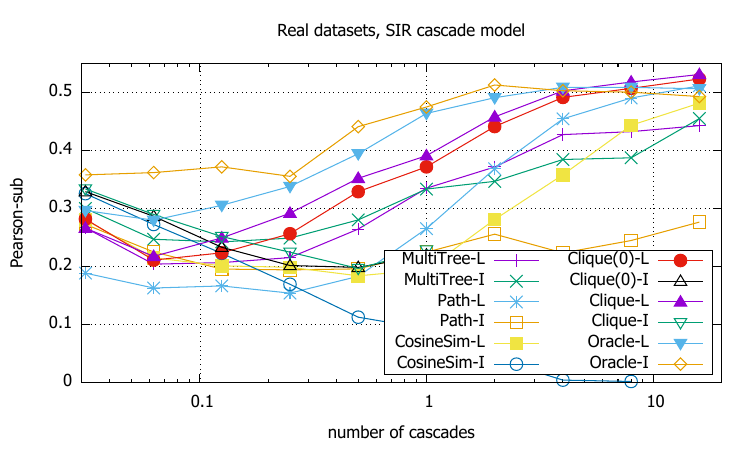}
\includegraphics[width=.49\textwidth]{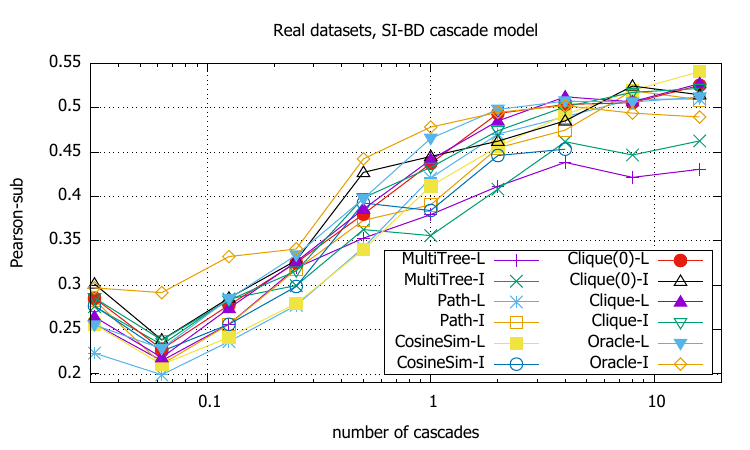}
\includegraphics[width=.49\textwidth]{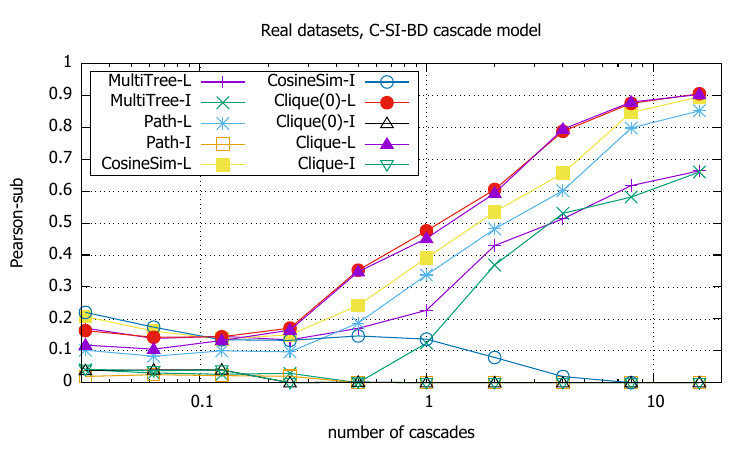}
\includegraphics[width=.49\textwidth]{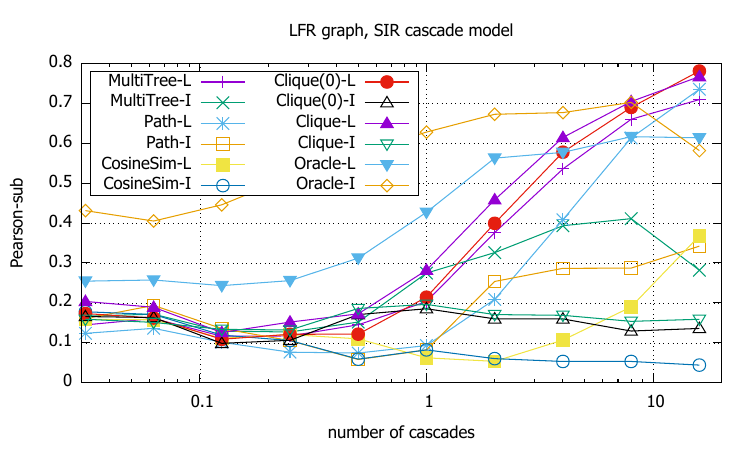}
\includegraphics[width=.49\textwidth]{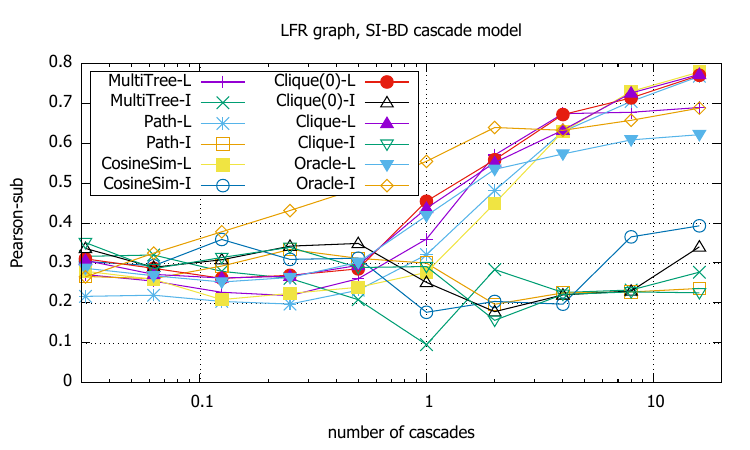}
\includegraphics[width=.49\textwidth]{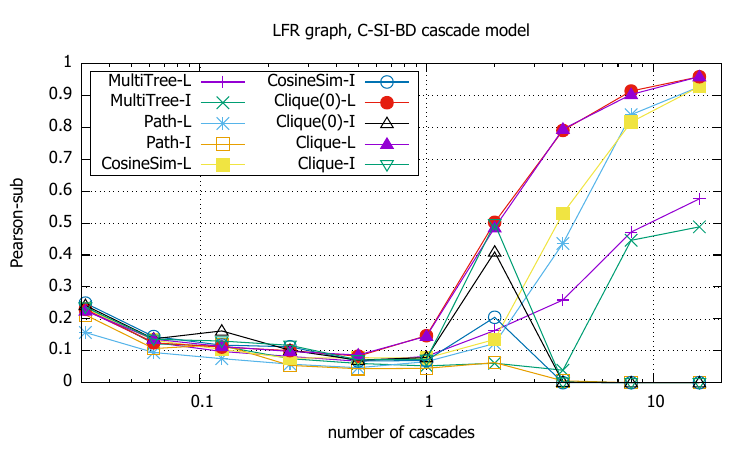}
\includegraphics[width=.49\textwidth]{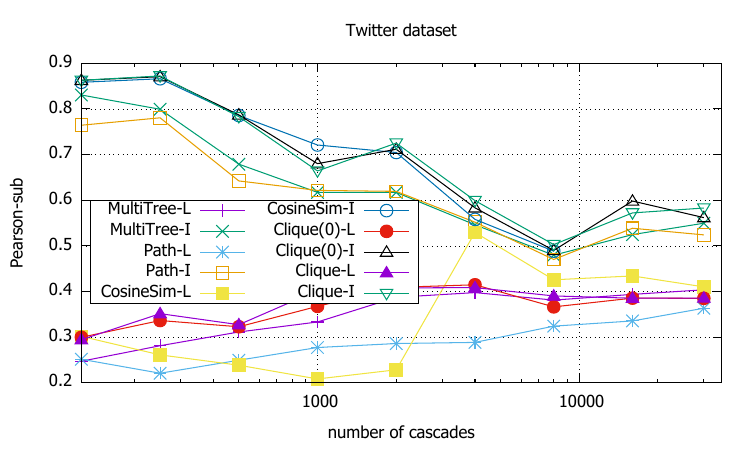}
\end{center}
\caption{Clustering of surrogate graphs with Louvain (``--L'') versus Infomap (``--I'').}
\label{fig:infomap}
\end{figure}

In this section, we analyze the effect of a clustering algorithm on the performance of approaches based on the clustering of surrogate graphs. In Figure~\ref{fig:infomap}, we compare Louvain with Infomap applied to \textsc{MultiTree}, \textsc{Path}, \textsc{CosineSim}, \textsc{Clique}, and \textsc{Oracle}. First, we observe that on the Twitter dataset, the results become extremely good, especially when the number of available cascades is small. However, the \textsc{Clique} algorithms are still among the best. In contrast, for other real datasets and for the synthetic LFR benchmark, the performance of Infomap is much worse, especially when the number of cascades is large. This holds for surrogate graphs, but not for \textsc{Oracle} and in some cases not for \textsc{MultiTree}. While analyzing this phenomenon, we noticed that for surrogate graphs, if the number of cascades is large, Infomap may predict degenerate clusterings consisting of only one cluster. This can be caused by the fact that surrogate graphs become dense when the number of cascades increases. In contrast to Infomap, Louvain is stable and works well even for dense graphs. On the other hand, on the Twitter dataset, Infomap significantly outperforms Louvain for a small number of cascades, i.e., when the density of surrogate graphs is not too large. This confirms that the choice of the clustering algorithm is essential. We conclude that Louvain leads to more stable results, but it is worth trying other options when possible.

\section{Conclusion}

In this article, we conducted a thorough analysis of inferring community structure based on cascade data. We compared algorithms recently proposed in the literature, simple heuristic approaches, and a more advanced approach based on mathematical modeling. Through a series of experiments on real-world and synthetic datasets, we conclude that the universal \textsc{Clique} method (combined with the Louvain clustering algorithm) is the best solution for the problem. This algorithm is extremely simple, efficient, and works well in all scenarios. In addition, this article covers several important methodological aspects specific to the problem at hand, e.g., choosing a performance measure to be used for comparing algorithms, aggregating results over multiple datasets, and generating realistic synthetic cascades. Another interesting conclusion is that standard clustering methods, such as the commonly used Louvain method, can be enhanced by running cascading processes on the network and weighing the node pairs based on the cascade data.

\begin{acks}
The work of Liudmila Prokhorenkova is supported by the Ministry of Education and Science of the Russian Federation in the framework of MegaGrant 075-15-2019-1926 and by the Russian President grant supporting leading scientific schools of the Russian Federation NSh-2540.2020.1.
\end{acks}

\bibliographystyle{ACM-Reference-Format}
\bibliography{acmart}

\appendix


\end{document}